\documentclass[pre,aps,amsmath,amssymb,reprint,floatfix,superscriptaddress,longbibliography]{revtex4-1}

\pdfoutput=1
\usepackage{xcolor}
\usepackage{siunitx}
\usepackage{graphicx}
\usepackage[colorlinks=true,citecolor=blue,urlcolor=blue]{hyperref}
\usepackage{etoolbox} 
\usepackage{bm} 
\usepackage[normalem]{ulem}

\def\cal#1{\mathcal{#1}}
\def\eqq#1{Eq.~(\ref{#1})}

\def\eq#1{(\ref{#1})}

\def\f#1{Fig.~\ref{#1}}

\def\c#1{~\cite{#1}}

\def\cc#1{~Ref.~\cite{#1}}

\def\av#1{\langle #1 \rangle}

\def\beq{\begin{equation}}
\def\eeq{\end{equation}}
\def\bea{\begin{eqnarray}}
\def\eea{\end{eqnarray}}

\def\kB{k_{\rm B}}
\def\tf{t_{\rm f}}
\def\kt{\kB T}

\def\e{{\rm e}}

\definecolor{blue}{rgb}{0,0,0}
\newcommand{\bb}[1]{\textcolor{blue}{#1}}

\begin{document}

\title{Increasing the clock speed of a thermodynamic computer by adding noise}

\author{Stephen Whitelam}
\email{swhitelam@lbl.gov}
\affiliation{Molecular Foundry, Lawrence Berkeley National Laboratory, 1 Cyclotron Road, Berkeley, CA 94720, USA}

\begin{abstract}
We describe a proposal for increasing the effective clock speed of a thermodynamic computer, by altering the interaction scale of the units within the computer and introducing to the computer an additional source of noise. The resulting thermodynamic computer program is equivalent to the original computer program, but runs at a higher clock speed. This approach offers a way of increasing the speed of thermodynamic computing while preserving the fidelity of computation.
\end{abstract}

\maketitle

{\em Introduction---} In classical computing, the energy scales of even the smallest devices, such as transistors and gates, are large compared to that of the thermal energy, $\kt$. As a result, there is a clear separation of scales between signal and noise, enabling deterministic computation. This determinism comes with an energy cost: classical computers operate far above the limits of thermodynamic efficiency, and require large amounts of power and heat dissipation to ensure their reliability\c{landauer1961irreversibility,bennett1982thermodynamics,ceruzzi2003history}. As devices shrink to the nanoscale, the energy associated with computation becomes comparable to that of thermal fluctuations, making computation more energy efficient\c{frank2017future,wolpert2019stochastic, conte2019thermodynamic}. However, the comparable scales of signal and noise makes deterministic computation challenging: efficient time-dependent protocols are required to do even the simplest logic operations on the $\kt$-scale\c{gingrich2016near,dago2021information,proesmans2020optimal,zulkowski2014optimal,proesmans2020finite,wimsatt2021harnessing,blaber2023optimal,barros2024learning}.

The field of thermodynamic computing views thermal fluctuations as a resource rather than a problem\c{conte2019thermodynamic,hylton2020thermodynamic}. Thermodynamic computing makes use of the tendency of physical systems to evolve toward thermal equilibrium to do computation. In this framework, the thermal bath contributes to computation by providing the fluctuations necessary for state changes, and in some modes of operation the signal {\em is} the noise, with the equilibrium fluctuations of the degrees of freedom of the thermodynamic computer being the output of the calculation\c{aifer2024thermodynamic, melanson2025thermodynamic}.
 \begin{figure*}[]
\centering
\includegraphics[width=\linewidth]{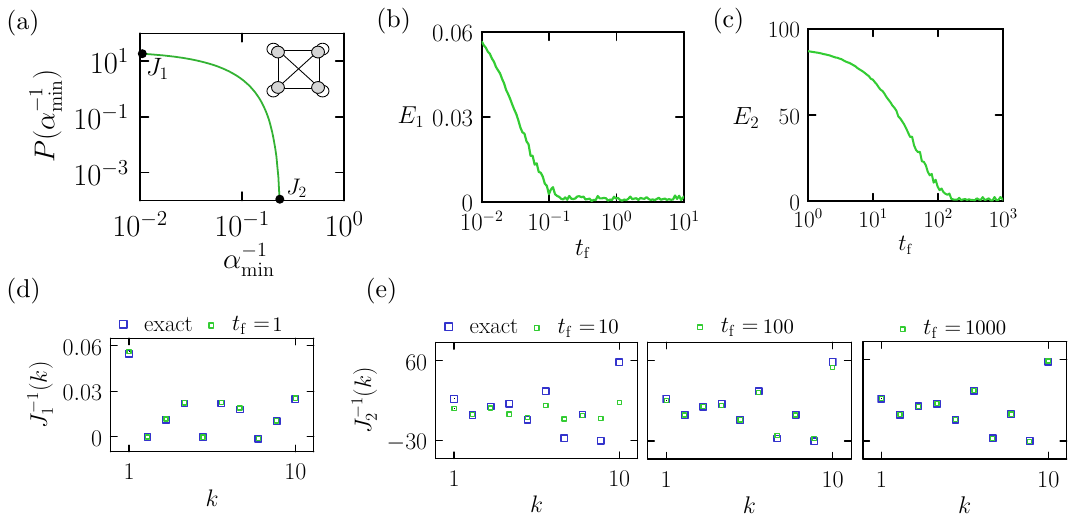}
\caption{Classical digital simulation of a thermodynamic computer program for matrix inversion\c{aifer2024thermodynamic, melanson2025thermodynamic}. (a) Probability distribution of the reciprocal of the smallest eigenvalue for $10^7$ $4 \times 4 $ symmetric positive definite matrices $J$. The values associated with the matrices $J_1$ and $J_2$ are marked by dots. Inset: schematic of the 4-unit thermodynamic computer used to estimate the inverses of $J_1$ and $J_2$, comprising 4 units and 10 connections. (b) Error $E_1$ [\eqq{frob}] in the estimate for $J_1^{-1}$ using the computer program \eq{lang1} run for time $\tf$; here and subsequently we state times in units of $\mu^{-1}$. The program is run $n_{\rm s}=10^4$ times, from which the averages $\av{S_i S_j}$ are calculated. (c) The same for the matrix $J_2^{-1}$. Note that the horizontal scales in (b) and (c) are different. (d) and (e): Estimates of the 10 distinct elements of $J_1^{-1}$ and $J_2^{-1}$, indexed by $k$, for the program times $\tf$ indicated. Note that the vertical axes in (d) and (e) have different scales.}
\label{fig1}
\end{figure*}

One challenge faced by thermodynamic computing is the requirement to attain thermodynamic equilibrium. Fields such as molecular self-assembly and glass physics reveal that although we may know the thermodynamic ground state of a physical system, there is no guarantee that a system will attain equilibrium on practically-realizable timescales\c{hagan2011mechanisms,whitelam2015statistical,biroli2013perspective,arceri2022glasses}. The emergence of slow dynamics and kinetic traps can cause a system to spend a considerable amount of time in a nonequilibrium state. With some thermodynamic computers\c{aifer2024thermodynamic, melanson2025thermodynamic} resembling spin glasses\c{mezard1984nature}, which can be very slow to equilibrate, strategies to attain equilibrium as rapidly as possible are important for realizing the potential of thermodynamic computing.

In this paper we propose a way to accelerate the dynamical evolution, and hence rate of equilibration, of the class of thermodynamic computers that evolve according to overdamped Langevin dynamics in the presence of Gaussian white noise. By altering the interaction scale of the units within the thermodynamic computer, and introducing to the system an additional source of Gaussian white noise whose variance is precisely chosen, we can effectively increase the clock speed of a thermodynamic computer with no impact on its computational ability. In other words, the efficiency of noise-dominated computing can be increased by {\em adding} noise. This counter-intuitive notion emphasizes a fundamental difference between thermodynamic computing and classical forms of computing (including analog computing) whose goal is to overcome or suppress noise.

In what follows we demonstrate this clock-acceleration procedure in general, using analytic methods. We provide a simple numerical illustration of the procedure, using a classical digital simulation of a thermodynamic computer program to invert a $4 \times 4$ symmetric positive definite matrix. The equilibration time of even this small system can vary by orders of magnitude as the matrix is changed. This broad distribution of timescales indicates the need for a clock-acceleration procedure, because it would be inconvenient for the run time of a thermodynamic computer program to vary by orders of magnitude as its parameters are varied.  

{\em Increasing the clock speed of a thermodynamic computer---} We consider a thermodynamic computer composed of $N$ units with real-valued scalar amplitudes $S_i$. Amplitudes could represent physical distances, if units are realized mechanically\c{dago2021information}, or voltage states, if realized by RLC circuits\c{aifer2024thermodynamic, melanson2025thermodynamic}, or phases, if realized by Josephson junctions\c{ray2023gigahertz}. The performance of the thermodynamic computer is then set by the characteristic time constants of these devices, roughly milliseconds, microseconds, or nanoseconds, respectively. Our simulation model runs on a conventional computer; when realized in {\em hardware}, the thermodynamic computer program runs automatically, driven only by thermal fluctuations\c{aifer2024thermodynamic, melanson2025thermodynamic}. Units interact via the potential $V({\bm S})$, which in existing thermodynamic computers consists of pairwise couplings between units\c{aifer2024thermodynamic, melanson2025thermodynamic}. 

Units evolve according to the overdamped Langevin dynamics
\beq
\label{lang1}
\dot{S}_i = -\mu \frac{\partial V({\bm S})}{\partial S_i}+ \sqrt{2 \mu \kt} \, \eta_i(t).
\eeq
Here $\mu$, the mobility, sets the time constant of the computer. For the thermodynamic computers described in Refs.\c{aifer2024thermodynamic,melanson2025thermodynamic}, which are realized by RLC circuits, $1/\mu \propto RC \sim 1$ microsecond. The combination $\kt$ is the thermal energy scale -- $k_{\rm B}$ is Boltzmann's constant and $T$ is temperature -- and $\eta$ is a Gaussian white noise with zero mean, $\av{\eta_i(t)}=0$, and covariance $\av{\eta_i(t) \eta_j(t')}=\delta_{ij} \delta(t-t')$. The Kronecker delta $\delta_{ij}$ indicates the independence of different noise components, and the Dirac delta $\delta(t-t')$ indicates the absence of time correlations. The noise $\eta$ represents the thermal fluctuations inherent to the system. We consider the interaction $V({\bm S})$ and the noise $\eta$ to specify the {\em program} of the thermodynamic computer.

Now say that we have a thermodynamic computer program \eq{lang1} whose dynamics is slow, and so takes a long time to run (and to equilibrate, if that is our goal). The most direct way to speed up the computer is to increase the mobility parameter $\mu$, which sets the computer's time constant. However, for RLC circuits we do not have unlimited freedom to do this, and timescales much shorter than microseconds are difficult to arrange.

An alternative is to consider running an accelerated computer program whose solution is identical to that of the original program but which takes less time to run. We can construct such a program as follows. Rescale the interaction term $V({\bm S})$ by a factor $\lambda \geq 1$ -- which can be done by uniformly rescaling the computer's coupling constants, no matter how high-order in ${\bm S}$ is the potential -- and introduce to the system an additional source of Gaussian white noise. Noise can be injected in a controlled way into thermodynamic computers\c{melanson2025thermodynamic}, single-molecule experiments~\c{chupeau2018thermal}, and information engines\c{saha2023information}. If the new noise has variance $\sigma^2$ then the new program is described by the equation
\beq
\label{lang3}
\dot{S}_i = -\mu \frac{\partial [\lambda V({\bm S})]}{\partial S_i}+ \sqrt{2 \mu \kt} \, \eta_i(t) + \sigma \zeta_i(t), 
\eeq
where $\av{\zeta_i(t)}=0$ and $\av{\zeta_i(t) \zeta_j(t')}=\delta_{ij} \delta(t-t')$. The added noise $\zeta$ can be thermal (from a different heat bath) or athermal in nature, provided that it is Gaussian and has no temporal correlations. The two noise terms in \eq{lang3} constitute an effective Gaussian white noise 
\beq
\sqrt{2 \mu \kt + \sigma^2} \, \xi_i(t),
\eeq 
where $\av{\xi_i(t)}=0$ and $\av{\xi_i(t) \xi_j(t')}=\delta_{ij} \delta(t-t')$. If we set 
\beq
\label{sigma}
\sigma^2 = 2 \mu \kt (\lambda -1),
\eeq
 then \eqq{lang3} can be written
\beq
\label{lang4}
\dot{S}_i = -\tilde{\mu} \frac{\partial V({\bm S})}{\partial S_i}+ \sqrt{2 \tilde{\mu} \kt} \, \xi_i(t),
\eeq
where $\tilde{\mu} \equiv \lambda \mu$. \eqq{lang4} is \eqq{lang1} with mobility parameter rescaled by a factor $\lambda \geq 1$ (the correlations of $\eta$ and $\xi$ are the same). Thus the accelerated computer program, \eq{lang3} and \eq{sigma}, runs the original computer program \eq{lang1} but faster~\footnote{A similar effect could be achieved by rescaling $V \to \lambda V$ and increasing temperature $T \to \lambda T$, but we assume scaling temperature is less practical than adding an external source of noise.}. 

\bb{Another way to understand this result is to rescale time $t \to \lambda t$ in \eq{lang1}: the result is \eqq{lang4}. Put another way, \eqq{lang4} can be written as \eqq{lang1} with a modified definition of time. That is, the thermodynamics of the modified program is {\em exactly} the same as the original program: the energy barriers may be higher, but the noise is proportionally larger, and so the modified system evolves across this landscape in exactly the way the original system would, and samples the same equilibrium distribution. The difference is that the modified system has a redefined scale of time, and reaches its destination sooner. In the language of computing, the accelerated program runs the original program with the clock speed increased by a factor of $\lambda \geq 1$.}

\bb{For this reason the clock-rescaling trick will also work with thermodynamic computers that do calculations out of equilibrium\c{whitelam2024thermodynamic}: the dynamical ensembles of \eq{lang1} and \eq{lang4} are identical, whether in or out of equilibrium. It will also work with nonlinear interactions $J_{1...N} S_1 ...S_N$, because under the rescaling $J_{1...N} \to \lambda J_{1...N}$ (and the addition of noise in the correct proportion) the modified Langevin equation looks like the original Langevin equation with a rescaled notion of time.}

{\em Numerical illustration of accelerated matrix inversion---} To illustrate this clock-acceleration procedure we consider a classical digital simulation of the matrix inversion program of Refs.\c{aifer2024thermodynamic,melanson2025thermodynamic}. Here the computer's degrees of freedom possess the bilinear pairwise interaction 
\beq
\label{nrg}
V({\bm S})=\sum_{\av{ij}} J_{ij} S_i S_j,
\eeq
where the sum runs over all distinct pairs of interactions. We choose an interaction matrix $J_{ij}$ that is symmetric and positive definite, and so has $N(N+1)/2$ distinct entries and all eigenvalues non-negative. This interaction is shown schematically in the inset of \f{fig1}(a). In thermal equilibrium the probability distribution of the computer's degrees of freedom is $\rho_0({\bm S}) = \e^{-\beta V({\bm S})}/\int {\rm d} {\bm S}'\e^{-\beta V({\bm S}')}$, where $\beta^{-1} \equiv \kt$, and so for the choice \eq{nrg} we have $\av{S_i}_0=0$ and
\beq
\label{inv}
\av{S_i S_j}_0 = \beta^{-1} (J^{-1})_{ij}.
\eeq
Here the brackets $\av{\cdot}_0$ denote a thermal average, and $(J^{-1})_{ij}$ denotes the elements of the inverse of the matrix $J$. Thus we can invert the matrix $J$ by sampling the thermal fluctuations, specifically the two-point correlations, of the units ${\bm S}$ in thermal equilibrium.

\bb{We consider the case $N = 4$. In  \f{fig1}(a) we show the probability distribution $P$ of the reciprocal of the smallest eigenvalue $\alpha_{\rm min}$ for $10^7$ symmetric positive-definite $4 \times 4$ matrices $J$. These matrices were generated by first constructing random symmetric matrices $B$, where each distinct lower-triangular element (including the diagonal) was drawn from a Gaussian distribution with zero mean and standard deviation 0.1. We then formed the matrix $A = B B^{\rm T}$, which is guaranteed to be symmetric and positive definite. To ensure that all pairwise interaction strengths in the thermodynamic computer were of order $k_{\rm B}T$ or larger, we normalized the resulting matrix by dividing all its elements by the magnitude of the smallest (in absolute value) element of $A$, yielding the matrix $\beta J$. This normalization step was arbitrary and used solely for convenience. The matrix $\beta J$ is guaranteed to be positive definite, and so all its eigenvalues are positive (note that individual elements of the matrix can still be negative).}

\bb{The reciprocal of the smallest eigenvalue of the matrix $J$ provides a rough estimate of the slowest relaxation timescale of the thermodynamic computer governed by \eqq{lang1}, when $J$ appears in \eq{nrg}. The relaxation of Langevin systems with a quadratic potential can be decomposed into modes corresponding to the eigenvectors of $J$. Each mode relaxes exponentially, with a rate proportional to the corresponding eigenvalue\c{gardiner2009stochastic}. The slowest mode therefore relaxes on a timescale inversely proportional to the smallest eigenvalue of $J$, making its reciprocal a useful proxy for the overall equilibration time.} We chose two matrices, $J_1$ and $J_2$, whose reciprocal smallest eigenvalues are 0.011 and 0.239, respectively. Based on this metric, we expect the program defined by $J_2$ to require at least an order of magnitude more time to equilibrate than the one defined by $J_1$. We shall see that this expectation is borne out in a qualitative sense, although the smallest eigenvalue is not a precise predictor of the equilibration time for two-point correlations.

The matrix $\beta J_1$ is~\footnote{For ease of display the matrix elements are shown to not more than 6 significant figures.}
\beq
\beta J_1 = \begin{pmatrix}
33.4484 & -2.13458 & -27.2213 & -13.3774 \\
-2.13458 & 91.0349 & 1 & 6.58694 \\
-27.2213 & 1 & 78.3708 & -12.075 \\
-13.3774 & 6.58694 & -12.075 & 55.602
\end{pmatrix}, \nonumber
\eeq
and its inverse is 
\begin{widetext}
\beq
\beta^{-1} J_1^{-1} = 10^{-3} \begin{pmatrix}
54.806 & -0.253332 & 21.8054 & 17.9513 \\
-0.253332 & 11.0905 & -0.456606 & -1.47396 \\
21.8054 & -0.456606 & 21.8886 & 10.0538 \\
17.9513 & -1.47396 & 10.0538 & 24.6619
\end{pmatrix}. \nonumber
\eeq
\end{widetext}
The matrix $\beta J_2$ is
\beq
\beta J_2 = \begin{pmatrix}
1.18752 & 1 & 1.00191 & 1.08641 \\
1 & 1.102484 & 1.02245 & 1.01064 \\
1.00191 & 1.02245 & 1.06828 & 1.03442 \\
1.08641 & 1.01064 & 1.03442 & 1.07344
\end{pmatrix}, \nonumber
\eeq
and its inverse is 
\beq
\beta^{-1} J_2^{-1} = \begin{pmatrix}
16.9091 & -1.08542 & 11.3872 & -27.0648 \\
-1.08542 & 8.37908 & -6.37593 & -0.646175 \\
11.3872 & -6.37593 & 25.497 & -30.092 \\
-27.0648 & -0.646175 & -30.092 & 57.9299
\end{pmatrix} \nonumber.
\eeq

We begin with the units of the thermodynamic computer set to zero, ${\bm S} = {\bm 0}$ (in thermal equilibrium we have $\av{{\bm S}}_0= {\bm 0}$). We run the program \eq{lang1} for time $\tf$, and then record the value of all distinct pairwise products $S_i S_j$. To carry out these simulations we used the first-order Euler discretization of \eq{lang1}, 
\beq
\label{lang2}
S_i(t+\Delta t) = S_i(t) -\mu \Delta t \frac{\partial V({\bm S})}{\partial S_i}+\sqrt{2 \mu \Delta t \kt}\, X,
\eeq
where $\Delta t=10^{-4} \mu^{-1}$ is the time step and $X \sim {\cal N}(0,1)$ is a Gaussian random number with zero mean and unit variance. \bb{We set $T = 1$ in numerical simulations, but retain $\beta$ in equations for clarity.}

We then reset the units of the computer to zero, and repeat the process, obtaining $n_{\rm s}=10^4$ sets of measurements in all. On a single device an alternative would be to run one long trajectory and sample periodically (note that the resetting procedure could be run in parallel, if many devices are available). Using all $n_{\rm s}= 10^4$ samples, we compute the average $\av{S_i S_j}$ of all distinct pairwise products. If $\tf$ is long enough for the computer to come to equilibrium, then $\av{S_i S_j} \approx \av{S_i S_j}_0$ (for sufficiently large $n_{\rm s}$), and these measurements will allow us to recover the inverse of the relevant matrix via \eqq{inv}.

As a measure of the error $E$ in the thermodynamic computer's estimate of the elements $\beta^{-1} (J^{-1})_{ij}$ we use the Frobenius norm
\beq
\label{frob}
E = \sqrt{\sum_{i,j} \left[ \langle S_i S_j \rangle - \beta^{-1} (J^{-1})_{ij} \right]^2}.
\eeq
 
 In \f{fig1}(b) we show the error $E_1$ in the estimate of the inverse $J_1^{-1}$, upon using the matrix $J_1$ in \eq{nrg} and running the program \eq{lang1} for time $\tf$ (in units of $\mu^{-1}$; subsequently, we will omit the units $\mu^{-1}$ when discussing timescales). For each run time, $n_{\rm s}=10^4$ samples are collected. The plot indicates that thermodynamic equilibrium is attained on timescales $\tf \gtrsim 0.2$.
 
 In \f{fig1}(c) we show the corresponding plot for the matrix $J_2$. In this case, equilibrium is attained for run times $\tf \gtrsim 200$, meaning that the run times for inverting the matrices $J_1$ and $J_2$ differ by about three orders of magnitude. The origin of this difference can be understood from the relative scales of the matrix elements involved: $J_1$ encodes a thermodynamic driving force of order 10s of $\kt$ (leading to rapid changes of ${\bm S}$), and the associated equilibrium fluctuations of the units ${\bm S}$ are on the scale of $10^{-2}$ (which take little time to establish). By contrast, $J_2$ encodes a thermodynamic driving force of order $\kt$ (leading to slower changes of ${\bm S}$), and some of the associated equilibrium fluctuations are of order 10 in magnitude (which take more time to establish).
 
 In \f{fig1}(d) and (e) we show the thermodynamic computer's estimate of the 10 distinct matrix elements (indexed by the variable $k$) of $J_1^{-1}$ (panel (d)) and $J_2^{-1}$ (panel (e)), for the indicated times, upon collecting $n_{\rm s}=10^4$ samples. An accurate estimate of $J_2^{-1}$ requires times in excess of 200, or of order 0.1 milliseconds for the clock speed of the thermodynamic computer of Refs.\c{aifer2024thermodynamic,melanson2025thermodynamic}. Given that $10^4$ samples are used, the total compute time would be of order a second, for a single $4 \times 4$ matrix (scaling estimates\c{aifer2024thermodynamic,melanson2025thermodynamic} suggest that the characteristic time for matrix inversion using an overdamped thermodynamic computer will grow as the square $N^2$ of the matrix dimension $N$, but matrices of equal size can have substantially different equilibration times).
 
 \begin{figure}[]
\centering
\includegraphics[width=\linewidth]{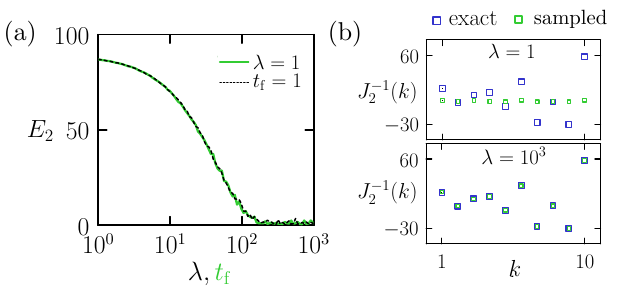}
\caption{Accelerated matrix inversion program. (a) Error $E_2$ [\eqq{frob}] in the estimate of the matrix $J_2^{-1}$ using the accelerated computer program \eq{lang3} and \eq{sigma} run for time $\tf=1$, as a function of the clock-acceleration parameter $\lambda$ (black dashed line). This result is overlaid on the data of \f{fig1}(c), derived from the original program \eq{lang1} run for time $\tf$ (green). We collect $n_{\rm s}=10^4$ samples for each program. As expected from the comparison of \eq{lang1} and \eq{lang4}, the accelerated program run for time $\tf$ is equivalent to the original program run for time $\lambda \tf$. (b) Exact elements of $J_2^{-1}$ and those estimated using the accelerated program run for time $\tf=1$ at two values of $\lambda$. The case $\lambda=1$ is equivalent to the original program.}
\label{fig2}
\end{figure}

To speed up the computation of $J_2^{-1}$ we can run the accelerated program specified by Eqns.~\eq{lang3} and \eq{sigma}. The matrix $J_2$ is therefore replaced by the matrix $\lambda J_2$, and we introduce an additional source of noise with variance \eq{sigma}. In \f{fig2}(a) we show the error $E_2$ in the computer's estimate of the elements of $J_2^{-1}$, as a function of the clock acceleration parameter $\lambda$. Here the program time is set to $\tf=1$, and $n_{\rm s}=10^4$ samples are collected. We overlay this result on that of \f{fig1}(c), the error associated with the original program \eq{lang1} run for time $\tf$. As expected from comparison of \eq{lang1} and \eq{lang4}, these results are essentially the same: the accelerated program run for $\tf$ is equivalent to the original program run for $\lambda \tf$.

Using the accelerated program with $\lambda=1000$, the total compute time for inverting $J_2$ at the clock speeds of the computer of Refs.\c {aifer2024thermodynamic,melanson2025thermodynamic} would be reduced from of order 1 second (for the original program) to of order 1 millisecond. If all samples were run on parallel hardware, the wall time for the computation would be of order 0.1 microseconds.

In panel (b) we show the inverse matrix elements extracted from the accelerated program \eq{lang3} and \eq{sigma} for two values of $\lambda$. As expected from the comparison of \eq{lang1} and \eq{lang4}, the elements extracted using $\lambda=1000$ (lower panel) for program time $\tf=1$ are as accurate as those extracted using the original program run for time $\tf=1000$ (right panel of \f{fig1}(e)).
 
{\em Conclusions---} As Moore's law runs its course, classical computing is approaching a local minimum of energy efficiency that is about 3 orders of magnitude removed from the thermodynamic limit of efficiency\c{hooker2021hardware,frank2017future,conte2019thermodynamic}. Quantum computing may supersede classical computing in the long term, but in the short term faces technical challenges to its large-scale deployment\c{preskill2018quantum}. Thermodynamic computing aims to bridge the gap between the two paradigms by providing new hardware and algorithms for energy-efficient computing on the $\kt$-scale\c{conte2019thermodynamic}. Its advantage over classical and quantum computing is that it uses thermal fluctuations rather than attempting to overcome or suppress them, but currently we possess relatively few examples of algorithms that can do computation in this fashion. Those that do exist have some promising characteristics. For instance, the thermodynamic matrix inversion algorithm can be run in a time that scales with matrix size less rapidly than do the corresponding classical algorithms\c{aifer2024thermodynamic}. However, the prefactors of these run times can vary considerably even between matrices of the same size, reflecting the general phenomenon that arbitrary physical systems can display a wide range of equilibration times.

Here we have proposed a way of accelerating the rate of thermodynamic computation by {\em adding} noise to the computer and scaling its inter-component connections appropriately. We have shown using classical digital simulation that a thermodynamic computer realized by overdamped Langevin dynamics in the presence of Gaussian white noise, described by \eqq{lang1}, can be accelerated by rescaling the interactions between its components and adding noise, according to Eqns.~\eq{lang3} and \eq{sigma}. The resulting dynamics, described by \eqq{lang4}, is equivalent to the original dynamics with the basic rate of computation or clock speed increased by a factor $\lambda \geq 1$. This acceleration of the clock leaves the dynamic and thermodynamic properties of the computer unchanged. We have illustrated this clock acceleration numerically, using the linear matrix inversion program of Refs.\c {aifer2024thermodynamic,melanson2025thermodynamic}. In general, the rescaling can be implemented for any nonlinear interaction $V({\bm S})$. In mathematical terms the $\lambda$-rescaling is a simple trick; its significance lies in the fact that it can be implemented directly in hardware, by uniformly rescaling the couplings of a thermodynamic computer and injecting Gaussian white noise with the required variance.

\bb{An important consideration for thermodynamic computing is the energy cost of performing calculations. In Appendix A we consider the addition of noise from the perspective of stochastic thermodynamics\c{sekimoto1998langevin,seifert2012stochastic}: added noise results in a power cost that scales as $P \propto \sigma^2$ (recall that $\sigma^2 \propto \kt  (\lambda -1)$. We can consider $\lambda$ to parameterize a family of programs, with larger values corresponding to faster computation and more energy expenditure, and smaller values corresponding to slower computation and less energy expenditure. For example, the time to solution of matrix inversion for an overdamped thermodynamic computer scales as $\kappa$, the condition number of the matrix\c{aifer2024thermodynamic}. If we wish to complete this calculation in time $\tf$, we require $\lambda = \kappa/\tf$, with associated energy cost $E \propto P \times \tf  \propto \kappa (1-\lambda^{-1}) \kt$. The cost of adding noise increases weakly with $\lambda$, because as $\lambda$ increases the time to solution decreases almost as rapidly as the energy cost increases.}

The \bb{theoretical} upper limit of $\lambda$ is set by the requirement that \eqq{lang1} remain valid. Specifically, the approximation that noise values at different times are statistically independent, $\av{\eta_i(t) \eta_i(t')} \propto \delta(t-t')$, holds as long as the computer's basic timescale $\mu^{-1}$ remains much larger than the autocorrelation time $\sim f^{-1}$ of the noise, where $\av{\eta_i(t) \eta_i(t')} \propto \e^{-f |t-t'|}$. If the injected noise is shot noise, then its autocorrelation time is set by the bandwidth $f$, and can approach 0.1 ns for electrical circuits\c{Horowitz2015}. This is about four orders of magnitude smaller than the microsecond time constants of the thermodynamic computers of Refs.\c{aifer2024thermodynamic,melanson2025thermodynamic}. On the basis of this comparison we can estimate an upper limit on $\lambda$ of about $10^3$, with values of 10 or 100 being conservative estimates for valid $\lambda$ values. A speed-up of a computer program of even an order of magnitude would represent a significant capability.

\bb{The global rescaling of the potential required by the clock-acceleration procedure is feasible in several physical implementations of thermodynamic computing. In RLC circuit-based computers\c{aifer2024thermodynamic,melanson2025thermodynamic}, where interactions are implemented as pairwise couplings through impedance or admittance networks, this rescaling can be achieved by uniformly adjusting the gains of amplifiers that set the coupling strengths, or by scaling all coupling elements (e.g., resistors or transconductance amplifiers) by a common factor. These modifications affect the potential landscape $V({\bm S})$ without altering the intrinsic dissipation rate of each node. The dissipation rate is governed by the local $RC$ time constant, which sets the mobility $\mu$. Likewise, in optical trap-based or mechanical implementations, interaction strengths can be uniformly scaled via laser intensity or spring stiffness\c{dago2021information}, respectively, without changing the damping constant or friction coefficient that sets the mobility.}

\bb{Controlled injection of additive Gaussian white noise with a specified variance is also experimentally feasible. In circuit-based implementations, this can be accomplished using dedicated noise generators or digital-to-analog converters fed with pseudorandom voltage signals\c{falcon2009fluctuations,melanson2025thermodynamic}. In optical trap systems, noise can be introduced by modulating the trap position using an acousto-optic deflector\c{gomez2010steady}. As long as the bandwidth of the injected noise is large compared to the characteristic timescale $\mu^{-1}$, the noise behaves as if it were delta-correlated.}

\bb{\eqq{lang1} assumes that the mobility $\mu$ is the same for all units. In the case of RLC circuit-based thermodynamic computers, this corresponds to all internal resistances $R_i$ being equal (see e.g. \f{fig1} of\cc{melanson2025thermodynamic}). This assumption does not constrain the computer's expressive power, because its computational ability is set by the programmable inter-unit couplings $V$, not by the individual mobilities of each unit.}

In this paper we have considered overdamped dynamics, but thermodynamic computers can also be realized by underdamped dynamics\c{aifer2024thermodynamic,melanson2025thermodynamic}, described generically by
\beq
\label{ud}
m\ddot{S}_i+ \gamma\dot{S}_i =- \frac{\partial V({\bm S})}{\partial S_i}+ \sqrt{2 \gamma \kt} \, \eta_i(t).
\eeq
The first term represents inertial effects ($m$ is mass), the second term is the damping force ($\gamma = 1/\mu$ is the damping coefficient), and the noise term represents thermal fluctuations inherent to the system. Rescaling time by a factor $\lambda \geq 1$ in \eq{ud} results in the equation
\beq
\label{ud2}
m\ddot{S}_i+ \lambda \gamma\dot{S}_i =- \lambda^2 \frac{\partial V({\bm S})}{\partial S_i}+ \sqrt{2 \lambda^3 \gamma \kt} \, \eta_i(t).
\eeq
We can produce \eq{ud2} from \eq{ud} -- and so effectively increase the clock speed of the computer -- by adding to \eq{ud} a Gaussian white noise with zero mean and variance $2 (\lambda^3 -1)\gamma \kt$, \bb{rescaling $V$ by a factor of $\lambda^2$, and rescaling the damping term by a factor of $\lambda$, assuming we have the parameters to do so.} Constructing an accelerated program for an underdamped thermodynamic computer would be more involved than doing so for an overdamped computer, but it is possible in principle.

{\em Acknowledgments ---} I thank David Sivak and Corneel Casert for discussions. This work was performed at the Molecular Foundry at Lawrence Berkeley National Laboratory, supported by the Office of Basic Energy Sciences of the U.S. Department of Energy under Contract No. DE-AC02--05CH11231.

\appendix

\section{\bb{Stochastic thermodynamics of added noise}}

The clock-rescaling trick described in the main text relies on the addition of a synthetic random noise to the existing thermal noise of the system. In this Appendix we develop a stochastic thermodynamics description of this added noise. Stochastic thermodynamics provides trajectory-level definitions of thermodynamic quantities such as heat, work, and entropy production. These definitions depend upon how forces are partitioned into conservative, dissipative, or external components\c{sekimoto1998langevin,seifert2012stochastic}.

In our case, while the microscopic dynamics and trajectory ensemble is uniquely specified by the equations of motion of the system, here \eqq{lang1}, the thermodynamic account depends in a qualitative sense on how we interpret the external noise. If we consider it to be a random force, the perspective that is natural from the standpoint of optical trap systems subject to random forcing\c{gomez2010steady}, then the system is out of equilibrium and produces entropy. If we consider the noise to contribute to the effective thermal bath, the perspective suggested by descriptions such as Eq. (5) of\cc{melanson2025thermodynamic}, then the system is in equilibrium with respect to an effective temperature parameter and produces no entropy.

The discussion in this Appendix emphasizes  how modeling choices affect the thermodynamic interpretation of a system. Several studies have explored this theme within stochastic thermodynamics. Studies of coarse graining emphasize that observational limitations can lead to underestimates of entropy production\c{esposito2012stochastic,shiraishi2015fluctuation}. In nonequilibrium systems, including active matter, non-thermal fluctuations and random forces can be incorporated into effective thermodynamic descriptions, or not, depending on the modeling framework\c{fodor2016how,marconi2017heat,puglisi2017temperature}. In several experimental settings it is now possible to add noise to a system in a controlled way, such as through random fluctuations of the power of a laser beam forming an optical trap, or by injecting shot noise into an electronic circuit\c{martinez2012effective,chupeau2018thermal,saha2023information,aifer2024thermodynamic}. We argue here that in such systems, the qualitative thermodynamic notion of whether a system is said to be in equilibrium or out of equilibrium, producing entropy or not, depends not only on the observed dynamics but also on the observer's assumptions about the origin of the system's fluctuations.

\subsection{Model}

Consider a simplified version of \eqq{lang1}, an overdamped Langevin equation for a single degree of freedom $x$,
\beq
\label{lang}
\gamma \dot{x}(t) = -\partial_x U(x,a) +  f(t)+ \xi(t).
\eeq
Here $\gamma=1/\mu$ is the damping coefficient, which we introduce so that each term in \eq{lang} has the dimensions of a force. The first term on the right-hand side of \eq{lang1} is a conservative force derived from a potential $U(x,a)$, where $a$ is a parameter of that potential. (In this Appendix we follow\cc{sekimoto1998langevin} and use $U$ rather than $V$ for the potential.) The term $f(t)$ is a Gaussian white noise with zero mean and correlation 
\beq
\langle f(t) f(t') \rangle = 2 \sigma^2 \delta(t - t'),
\eeq
while $\xi(t)$ is a Gaussian white noise with zero mean and correlation
\beq
\langle \xi(t) \xi(t') \rangle = 2\gamma k_B T \delta(t - t').
\eeq
The term $\xi(t)$ models fluctuations from a thermal bath at temperature $T$, while $f(t)$ (called $\zeta_i(t)$ in the main text) describes an additional random noise. The two noise sources are uncorrelated, with $\av{f(t) \xi(t)}=0$. In a thermodynamic computer, the added noise is electronic in nature. For a colloidal particle confined in an optical trap, with $U(x,a) = k(x-a)^2/2$\c{Ciliberto2017experiments} the added noise could arise from various sources, such as mechanical shaking of the sample stage, fluctuating laser intensity, or feedback from an experimental control system\c{gomez2010steady,martinez2012effective,chupeau2018thermal,saha2023information}.

\subsection{The force picture}
\begin{table*}
\centering
\begin{tabular}{c|c|c}
\textbf{quantity} & \textbf{force picture} & \textbf{bath picture} \\
\hline
origin of \(f(t)\) & external force & effective thermal environment \\
heat flow & \(dQ = -(-\gamma \dot{x} + \xi(t)) \circ dx\) & \(d\tilde{Q} = -(-\gamma \dot{x} + f(t) + \xi(t)) \circ dx\) \\
work & \(dW = f(t) \circ dx\) & $d \tilde{W} =0$ \\
entropy production rate & $\dot{S} = \sigma^2 / (\gamma k_B T)$ &  $\dot{\tilde{S}}= 0$ \\
time-reversal symmetry & broken & preserved \\
status of system & nonequilibrium & equilibrium \\
\hline
\end{tabular}
\caption{Comparison of the force- and bath interpretations of the Langevin equation \eq{lang}, for constant potential parameter $a$.}
\label{tab1}
\end{table*}

Following\cc{sekimoto1998langevin}, we can collect the terms of \eqq{lang} into physically suggestive groups, and multiply by $dx$ to examine the energy balance upon a change of state of the system. One such grouping yields the equation
\beq
\label{nrg2}
\underbrace{f(t) \circ dx+ \frac{\partial U}{\partial a} \circ da}_{dW}=\underbrace{ -(-\gamma \dot{x} + \xi (t)) \circ dx}_{dQ} + dU,
\eeq
where we have introduced the total differential of the potential,
\beq
dU = \frac{\partial U}{\partial x} \circ dx + \frac{\partial U}{\partial a} \circ da,
\eeq
and the symbol $\circ$ denotes the Stratonovich product. 

\eqq{nrg2} defines what we shall refer to as the {\em force picture}, because it views $f(t)$ as an external force. The left-hand side of \eq{nrg2} represents the work $d W$ done on the system, while the first term on the right-hand side is the heat $d Q$ transferred from the system to the bath (the sign convention is that of\cc{sekimoto1998langevin}). The second term on the right-hand side of \eq{nrg2} is the change $d U$ of internal energy. \eqq{nrg2} is a representation of the first law of thermodynamics at the trajectory level.

In the force picture, $f(t)$ does work on the system. If we do not change the potential parameter $a$, then the mean rate of work, $\dot{W} = \av{f(t) \circ \dot{x}}$, is
\bea
\dot{W} &=& \left\langle f(t) \circ \dot{x}(t) \right\rangle \nonumber \\
&=&\left \langle f(t) \circ \gamma^{-1} \left( - \partial_x U(x,a) + f(t)+ \xi(t) \right)\right\rangle \nonumber \\
&=&\gamma^{-1} \left\langle f(t) \circ f(t) \right\rangle \nonumber \\
&=& \gamma^{-1} \sigma^2.
\eea
Once the system has reached steady state, this work must be continually dissipated as heat, and so the system must produce entropy (relative to the thermal bath at $T$) at a rate
\beq
\dot{S} =  \frac{\sigma^2}{\gamma \kt}.
\eeq

The entropy production rate is positive for $\sigma^2 > 0$, and represents the irreversible dissipation of the energy injected by the random forcing. Even though $f(t)$ is stochastic, it represents an external drive that breaks time-reversal symmetry, and sustains a nonequilibrium steady state.

\subsection{The bath picture}

However, we can construct an alternative thermodynamic interpretation of \eqq{lang}. Returning to \eq{nrg2}, we take the term $f(t) \circ dx$ to the right-hand side to obtain
\bea
\label{bath}
\underbrace{ \frac{\partial U}{\partial a} \circ da}_{d\tilde{W}}&=& -(-\gamma \dot{x} + f(t)+ \xi (t)) \circ dx + dU \nonumber \\
&\equiv&\underbrace{ -(-\gamma \dot{x} + \tilde{\xi}(t)) \circ dx}_{d\tilde{Q}} + dU. \nonumber
\eea
Work and heat are now defined by the quantities $d \tilde{W}$ and $d\tilde{Q}$. In this picture, called the {\em bath picture}, we have absorbed $f(t)$ into an effective Gaussian white noise $\tilde{\xi}(t) \equiv \xi(t)+f(t)$, whose correlations satisfy $\av{\tilde{\xi}(t)}=0$ and 
\beq
\av{\tilde{\xi}(t) \tilde{\xi}(t')}=2 \gamma \kt + 2\sigma^2 \equiv 2 \gamma  k_{\rm B} \tilde{T}.
\eeq
This relation defines an effective temperature $\tilde{T}=T+\sigma^2/(\gamma k_{\rm B})$. If we do not change the potential parameter $a$ then there is no external force acting on the system, and so the entropy production rate $\dot{\tilde{S}}$ is zero~\footnote{Explicitly, we can show from \eqq{action} that the heat flow vanishes, $q[x(t)]=\tilde{T}[\mathcal{A}(x(t),0) - \mathcal{A}(x(\tf-t),0)]=0$, and so therefore does the entropy production.}. The system is in equilibrium with respect to a thermal bath at effective temperature $\tilde{T}$. 

\subsection{Discussion}

In stochastic thermodynamics, thermodynamic structure depends both on the form of stochastic equation and on how forces are partitioned\c{sekimoto1998langevin,seifert2012stochastic}. Here we have focused on this partitioning in the context of added random noise, where different physical interpretations of the same microscopic dynamics give rise to qualitatively different thermodynamic narratives. 

The thermodynamic implications of the force and bath picture are summarized (for the case in which the potential parameter $a$ is constant) in Table~\ref{tab1}. Different assumptions about the origin of the terms in \eqq{lang} leads to different thermodynamic pictures. In the force picture, the random fluctuation $f(t)$ is assumed to be an external force that performs work on the system. This work is dissipated into the thermal environment, and so the steady-state rate of entropy production in the medium is nonzero. By contrast, in the bath picture the same dynamics is described by a system in contact with an effective thermal bath at temperature $\tilde{T} = T + \sigma^2 /(\gamma  k_{\rm B})$. In this picture, there is no net flow of energy and no entropy production. The observed fluctuations are attributed entirely to equilibrium thermal agitation at elevated temperature. 

For the clock-rescaling procedure described in the main text, the rescaling of the potential affects, numerically, the heat emitted and the work associated with changes of the potential. The addition of noise, however, allows for two {\em qualitatively} distinct thermodynamic descriptions of the system. Our discussion emphasizes that even qualitative thermodynamic judgments, such as whether a system is dissipative or in equilibrium, depend fundamentally on how an observer interprets the forces acting on a system. 

It is important to stress that the trajectory ensemble associated with the force and bath pictures is an invariant physical observable. It is most convenient to calculate this within the bath picture, where it can be seen that \eqq{lang} generates trajectories $x(t)$ with a weight\c{seifert2012stochastic}
\beq
p[x(t)]  \propto \exp\left[-\mathcal{A}(x(t), a(t))\right].
\eeq
Here
\beq
\label{action}
\mathcal{A}(x(t), a(t)) \equiv \int_0^{\tf} {\rm d} t \left[ \frac{(\gamma \dot{x} -\partial_x U)^2}{4\tilde{D}} + \frac{1}{2}\partial_x^2 U \right]
\eeq
is the action associated with the trajectory, and $\tilde{D} \equiv  \gamma \kt + \sigma^2$.

We end by noting that the distinction between the force and bath pictures does not need to be binary. We could in principle assign an arbitrary portion $(1-\theta) f(t)$ of the noise $f(t)$ to the bath, and the remainder $\theta f(t)$ to external forcing, where $\theta \in [0,1]$. This freedom defines a continuum of thermodynamically distinct but dynamically equivalent models, parameterized by a variable that controls the partitioning of noise. In this respect, we can consider each value of $\theta$ to define a distinct thermodynamic ``gauge''. Each gauge preserves the physical observable, the underlying stochastic dynamics, but possesses a distinct notion of work, entropy production, and effective temperature.


\begin{thebibliography}{44}%
\makeatletter
\providecommand \@ifxundefined [1]{%
 \@ifx{#1\undefined}
}%
\providecommand \@ifnum [1]{%
 \ifnum #1\expandafter \@firstoftwo
 \else \expandafter \@secondoftwo
 \fi
}%
\providecommand \@ifx [1]{%
 \ifx #1\expandafter \@firstoftwo
 \else \expandafter \@secondoftwo
 \fi
}%
\providecommand \natexlab [1]{#1}%
\providecommand \enquote  [1]{``#1''}%
\providecommand \bibnamefont  [1]{#1}%
\providecommand \bibfnamefont [1]{#1}%
\providecommand \citenamefont [1]{#1}%
\providecommand \href@noop [0]{\@secondoftwo}%
\providecommand \href [0]{\begingroup \@sanitize@url \@href}%
\providecommand \@href[1]{\@@startlink{#1}\@@href}%
\providecommand \@@href[1]{\endgroup#1\@@endlink}%
\providecommand \@sanitize@url [0]{\catcode `\\12\catcode `\$12\catcode
  `\&12\catcode `\#12\catcode `\^12\catcode `\_12\catcode `\%12\relax}%
\providecommand \@@startlink[1]{}%
\providecommand \@@endlink[0]{}%
\providecommand \url  [0]{\begingroup\@sanitize@url \@url }%
\providecommand \@url [1]{\endgroup\@href {#1}{\urlprefix }}%
\providecommand \urlprefix  [0]{URL }%
\providecommand \Eprint [0]{\href }%
\providecommand \doibase [0]{http://dx.doi.org/}%
\providecommand \selectlanguage [0]{\@gobble}%
\providecommand \bibinfo  [0]{\@secondoftwo}%
\providecommand \bibfield  [0]{\@secondoftwo}%
\providecommand \translation [1]{[#1]}%
\providecommand \BibitemOpen [0]{}%
\providecommand \bibitemStop [0]{}%
\providecommand \bibitemNoStop [0]{.\EOS\space}%
\providecommand \EOS [0]{\spacefactor3000\relax}%
\providecommand \BibitemShut  [1]{\csname bibitem#1\endcsname}%
\let\auto@bib@innerbib\@empty
\bibitem [{\citenamefont {Landauer}(1961)}]{landauer1961irreversibility}%
  \BibitemOpen
  \bibfield  {author} {\bibinfo {author} {\bibfnamefont {Rolf}\ \bibnamefont
  {Landauer}},\ }\bibfield  {title} {\enquote {\bibinfo {title}
  {Irreversibility and heat generation in the computing process},}\ }\href@noop
  {} {\bibfield  {journal} {\bibinfo  {journal} {IBM journal of research and
  development}\ }\textbf {\bibinfo {volume} {5}},\ \bibinfo {pages} {183--191}
  (\bibinfo {year} {1961})}\BibitemShut {NoStop}%
\bibitem [{\citenamefont {Bennett}(1982)}]{bennett1982thermodynamics}%
  \BibitemOpen
  \bibfield  {author} {\bibinfo {author} {\bibfnamefont {Charles~H.}\
  \bibnamefont {Bennett}},\ }\bibfield  {title} {\enquote {\bibinfo {title}
  {The thermodynamics of computation--a review},}\ }\href@noop {} {\bibfield
  {journal} {\bibinfo  {journal} {International Journal of Theoretical
  Physics}\ }\textbf {\bibinfo {volume} {21}},\ \bibinfo {pages} {905--940}
  (\bibinfo {year} {1982})}\BibitemShut {NoStop}%
\bibitem [{\citenamefont {Ceruzzi}(2003)}]{ceruzzi2003history}%
  \BibitemOpen
  \bibfield  {author} {\bibinfo {author} {\bibfnamefont {Paul~E.}\ \bibnamefont
  {Ceruzzi}},\ }\href@noop {} {\emph {\bibinfo {title} {A history of modern
  computing}}}\ (\bibinfo  {publisher} {MIT Press},\ \bibinfo {year}
  {2003})\BibitemShut {NoStop}%
\bibitem [{\citenamefont {Frank}(2017)}]{frank2017future}%
  \BibitemOpen
  \bibfield  {author} {\bibinfo {author} {\bibfnamefont {Michael~P}\
  \bibnamefont {Frank}},\ }\bibfield  {title} {\enquote {\bibinfo {title} {The
  future of computing depends on making it reversible},}\ }\href@noop {}
  {\bibfield  {journal} {\bibinfo  {journal} {IEEE Spectrum}\ }\textbf
  {\bibinfo {volume} {25}},\ \bibinfo {pages} {2017} (\bibinfo {year}
  {2017})}\BibitemShut {NoStop}%
\bibitem [{\citenamefont {Wolpert}(2019)}]{wolpert2019stochastic}%
  \BibitemOpen
  \bibfield  {author} {\bibinfo {author} {\bibfnamefont {David~H.}\
  \bibnamefont {Wolpert}},\ }\bibfield  {title} {\enquote {\bibinfo {title}
  {The stochastic thermodynamics of computation},}\ }\href@noop {} {\bibfield
  {journal} {\bibinfo  {journal} {Journal of Physics A: Mathematical and
  Theoretical}\ }\textbf {\bibinfo {volume} {52}},\ \bibinfo {pages} {193001}
  (\bibinfo {year} {2019})}\BibitemShut {NoStop}%
\bibitem [{\citenamefont {Conte}\ \emph {et~al.}(2019)\citenamefont {Conte},
  \citenamefont {DeBenedictis}, \citenamefont {Ganesh}, \citenamefont {Hylton},
  \citenamefont {Crooks}, \citenamefont {Coles} \emph
  {et~al.}}]{conte2019thermodynamic}%
  \BibitemOpen
  \bibfield  {author} {\bibinfo {author} {\bibfnamefont {Tom}\ \bibnamefont
  {Conte}}, \bibinfo {author} {\bibfnamefont {Erik}\ \bibnamefont
  {DeBenedictis}}, \bibinfo {author} {\bibfnamefont {Natesh}\ \bibnamefont
  {Ganesh}}, \bibinfo {author} {\bibfnamefont {Todd}\ \bibnamefont {Hylton}},
  \bibinfo {author} {\bibfnamefont {Gavin~E.}\ \bibnamefont {Crooks}}, \bibinfo
  {author} {\bibfnamefont {Patrick~J.}\ \bibnamefont {Coles}},  \emph
  {et~al.},\ }\bibfield  {title} {\enquote {\bibinfo {title} {Thermodynamic
  computing},}\ }\href@noop {} {\bibfield  {journal} {\bibinfo  {journal}
  {arXiv preprint arXiv:1911.01968}\ } (\bibinfo {year} {2019})}\BibitemShut
  {NoStop}%
\bibitem [{\citenamefont {Gingrich}\ \emph {et~al.}(2016)\citenamefont
  {Gingrich}, \citenamefont {Rotskoff}, \citenamefont {Crooks},\ and\
  \citenamefont {Geissler}}]{gingrich2016near}%
  \BibitemOpen
  \bibfield  {author} {\bibinfo {author} {\bibfnamefont {Todd~R}\ \bibnamefont
  {Gingrich}}, \bibinfo {author} {\bibfnamefont {Grant~M}\ \bibnamefont
  {Rotskoff}}, \bibinfo {author} {\bibfnamefont {Gavin~E}\ \bibnamefont
  {Crooks}}, \ and\ \bibinfo {author} {\bibfnamefont {Phillip~L}\ \bibnamefont
  {Geissler}},\ }\bibfield  {title} {\enquote {\bibinfo {title} {Near-optimal
  protocols in complex nonequilibrium transformations},}\ }\href@noop {}
  {\bibfield  {journal} {\bibinfo  {journal} {Proceedings of the National
  Academy of Sciences}\ }\textbf {\bibinfo {volume} {113}},\ \bibinfo {pages}
  {10263--10268} (\bibinfo {year} {2016})}\BibitemShut {NoStop}%
\bibitem [{\citenamefont {Dago}\ \emph {et~al.}(2021)\citenamefont {Dago},
  \citenamefont {Pereda}, \citenamefont {Barros}, \citenamefont {Ciliberto},\
  and\ \citenamefont {Bellon}}]{dago2021information}%
  \BibitemOpen
  \bibfield  {author} {\bibinfo {author} {\bibfnamefont {Salambo}\ \bibnamefont
  {Dago}}, \bibinfo {author} {\bibfnamefont {Jorge}\ \bibnamefont {Pereda}},
  \bibinfo {author} {\bibfnamefont {Nicolas}\ \bibnamefont {Barros}}, \bibinfo
  {author} {\bibfnamefont {Sergio}\ \bibnamefont {Ciliberto}}, \ and\ \bibinfo
  {author} {\bibfnamefont {Ludovic}\ \bibnamefont {Bellon}},\ }\bibfield
  {title} {\enquote {\bibinfo {title} {Information and thermodynamics: fast and
  precise approach to {L}andauer's bound in an underdamped micromechanical
  oscillator},}\ }\href@noop {} {\bibfield  {journal} {\bibinfo  {journal}
  {Physical Review Letters}\ }\textbf {\bibinfo {volume} {126}},\ \bibinfo
  {pages} {170601} (\bibinfo {year} {2021})}\BibitemShut {NoStop}%
\bibitem [{\citenamefont {Proesmans}\ \emph
  {et~al.}(2020{\natexlab{a}})\citenamefont {Proesmans}, \citenamefont
  {Ehrich},\ and\ \citenamefont {Bechhoefer}}]{proesmans2020optimal}%
  \BibitemOpen
  \bibfield  {author} {\bibinfo {author} {\bibfnamefont {Karel}\ \bibnamefont
  {Proesmans}}, \bibinfo {author} {\bibfnamefont {Jannik}\ \bibnamefont
  {Ehrich}}, \ and\ \bibinfo {author} {\bibfnamefont {John}\ \bibnamefont
  {Bechhoefer}},\ }\bibfield  {title} {\enquote {\bibinfo {title} {Optimal
  finite-time bit erasure under full control},}\ }\href@noop {} {\bibfield
  {journal} {\bibinfo  {journal} {Physical Review E}\ }\textbf {\bibinfo
  {volume} {102}},\ \bibinfo {pages} {032105} (\bibinfo {year}
  {2020}{\natexlab{a}})}\BibitemShut {NoStop}%
\bibitem [{\citenamefont {Zulkowski}\ and\ \citenamefont
  {DeWeese}(2014)}]{zulkowski2014optimal}%
  \BibitemOpen
  \bibfield  {author} {\bibinfo {author} {\bibfnamefont {Patrick~R}\
  \bibnamefont {Zulkowski}}\ and\ \bibinfo {author} {\bibfnamefont {Michael~R}\
  \bibnamefont {DeWeese}},\ }\bibfield  {title} {\enquote {\bibinfo {title}
  {Optimal finite-time erasure of a classical bit},}\ }\href@noop {} {\bibfield
   {journal} {\bibinfo  {journal} {Physical Review E}\ }\textbf {\bibinfo
  {volume} {89}},\ \bibinfo {pages} {052140} (\bibinfo {year}
  {2014})}\BibitemShut {NoStop}%
\bibitem [{\citenamefont {Proesmans}\ \emph
  {et~al.}(2020{\natexlab{b}})\citenamefont {Proesmans}, \citenamefont
  {Ehrich},\ and\ \citenamefont {Bechhoefer}}]{proesmans2020finite}%
  \BibitemOpen
  \bibfield  {author} {\bibinfo {author} {\bibfnamefont {Karel}\ \bibnamefont
  {Proesmans}}, \bibinfo {author} {\bibfnamefont {Jannik}\ \bibnamefont
  {Ehrich}}, \ and\ \bibinfo {author} {\bibfnamefont {John}\ \bibnamefont
  {Bechhoefer}},\ }\bibfield  {title} {\enquote {\bibinfo {title} {Finite-time
  {L}andauer principle},}\ }\href@noop {} {\bibfield  {journal} {\bibinfo
  {journal} {Physical Review Letters}\ }\textbf {\bibinfo {volume} {125}},\
  \bibinfo {pages} {100602} (\bibinfo {year} {2020}{\natexlab{b}})}\BibitemShut
  {NoStop}%
\bibitem [{\citenamefont {Wimsatt}\ \emph {et~al.}(2021)\citenamefont
  {Wimsatt}, \citenamefont {Saira}, \citenamefont {Boyd}, \citenamefont
  {Matheny}, \citenamefont {Han}, \citenamefont {Roukes},\ and\ \citenamefont
  {Crutchfield}}]{wimsatt2021harnessing}%
  \BibitemOpen
  \bibfield  {author} {\bibinfo {author} {\bibfnamefont {Gregory}\ \bibnamefont
  {Wimsatt}}, \bibinfo {author} {\bibfnamefont {Olli-Pentti}\ \bibnamefont
  {Saira}}, \bibinfo {author} {\bibfnamefont {Alexander~B}\ \bibnamefont
  {Boyd}}, \bibinfo {author} {\bibfnamefont {Matthew~H}\ \bibnamefont
  {Matheny}}, \bibinfo {author} {\bibfnamefont {Siyuan}\ \bibnamefont {Han}},
  \bibinfo {author} {\bibfnamefont {Michael~L}\ \bibnamefont {Roukes}}, \ and\
  \bibinfo {author} {\bibfnamefont {James~P}\ \bibnamefont {Crutchfield}},\
  }\bibfield  {title} {\enquote {\bibinfo {title} {Harnessing fluctuations in
  thermodynamic computing via time-reversal symmetries},}\ }\href@noop {}
  {\bibfield  {journal} {\bibinfo  {journal} {Physical Review Research}\
  }\textbf {\bibinfo {volume} {3}},\ \bibinfo {pages} {033115} (\bibinfo {year}
  {2021})}\BibitemShut {NoStop}%
\bibitem [{\citenamefont {Blaber}\ and\ \citenamefont
  {Sivak}(2023)}]{blaber2023optimal}%
  \BibitemOpen
  \bibfield  {author} {\bibinfo {author} {\bibfnamefont {Steven}\ \bibnamefont
  {Blaber}}\ and\ \bibinfo {author} {\bibfnamefont {David~A}\ \bibnamefont
  {Sivak}},\ }\bibfield  {title} {\enquote {\bibinfo {title} {Optimal control
  in stochastic thermodynamics},}\ }\href@noop {} {\bibfield  {journal}
  {\bibinfo  {journal} {Journal of Physics Communications}\ }\textbf {\bibinfo
  {volume} {7}},\ \bibinfo {pages} {033001} (\bibinfo {year}
  {2023})}\BibitemShut {NoStop}%
\bibitem [{\citenamefont {Barros}\ \emph {et~al.}(2024)\citenamefont {Barros},
  \citenamefont {Whitelam}, \citenamefont {Ciliberto},\ and\ \citenamefont
  {Bellon}}]{barros2024learning}%
  \BibitemOpen
  \bibfield  {author} {\bibinfo {author} {\bibfnamefont {Nicolas}\ \bibnamefont
  {Barros}}, \bibinfo {author} {\bibfnamefont {Stephen}\ \bibnamefont
  {Whitelam}}, \bibinfo {author} {\bibfnamefont {Sergio}\ \bibnamefont
  {Ciliberto}}, \ and\ \bibinfo {author} {\bibfnamefont {Ludovic}\ \bibnamefont
  {Bellon}},\ }\bibfield  {title} {\enquote {\bibinfo {title} {Learning
  efficient erasure protocols for an underdamped memory},}\ }\href@noop {}
  {\bibfield  {journal} {\bibinfo  {journal} {arXiv preprint arXiv:2409.15050}\
  } (\bibinfo {year} {2024})}\BibitemShut {NoStop}%
\bibitem [{\citenamefont {Hylton}(2020)}]{hylton2020thermodynamic}%
  \BibitemOpen
  \bibfield  {author} {\bibinfo {author} {\bibfnamefont {Todd}\ \bibnamefont
  {Hylton}},\ }\bibfield  {title} {\enquote {\bibinfo {title} {Thermodynamic
  computing: an intellectual and technological frontier},}\ }in\ \href@noop {}
  {\emph {\bibinfo {booktitle} {Proceedings}}},\ Vol.~\bibinfo {volume} {47}\
  (\bibinfo {organization} {MDPI},\ \bibinfo {year} {2020})\ p.~\bibinfo
  {pages} {23}\BibitemShut {NoStop}%
\bibitem [{\citenamefont {Aifer}\ \emph {et~al.}(2024)\citenamefont {Aifer},
  \citenamefont {Donatella}, \citenamefont {Gordon}, \citenamefont {Duffield},
  \citenamefont {Ahle}, \citenamefont {Simpson}, \citenamefont {Crooks},\ and\
  \citenamefont {Coles}}]{aifer2024thermodynamic}%
  \BibitemOpen
  \bibfield  {author} {\bibinfo {author} {\bibfnamefont {Maxwell}\ \bibnamefont
  {Aifer}}, \bibinfo {author} {\bibfnamefont {Kaelan}\ \bibnamefont
  {Donatella}}, \bibinfo {author} {\bibfnamefont {Max~Hunter}\ \bibnamefont
  {Gordon}}, \bibinfo {author} {\bibfnamefont {Samuel}\ \bibnamefont
  {Duffield}}, \bibinfo {author} {\bibfnamefont {Thomas}\ \bibnamefont {Ahle}},
  \bibinfo {author} {\bibfnamefont {Daniel}\ \bibnamefont {Simpson}}, \bibinfo
  {author} {\bibfnamefont {Gavin}\ \bibnamefont {Crooks}}, \ and\ \bibinfo
  {author} {\bibfnamefont {Patrick~J}\ \bibnamefont {Coles}},\ }\bibfield
  {title} {\enquote {\bibinfo {title} {Thermodynamic linear algebra},}\
  }\href@noop {} {\bibfield  {journal} {\bibinfo  {journal} {npj Unconventional
  Computing}\ }\textbf {\bibinfo {volume} {1}},\ \bibinfo {pages} {13}
  (\bibinfo {year} {2024})}\BibitemShut {NoStop}%
\bibitem [{\citenamefont {Melanson}\ \emph {et~al.}(2025)\citenamefont
  {Melanson}, \citenamefont {Abu~Khater}, \citenamefont {Aifer}, \citenamefont
  {Donatella}, \citenamefont {Hunter~Gordon}, \citenamefont {Ahle},
  \citenamefont {Crooks}, \citenamefont {Martinez}, \citenamefont {Sbahi},\
  and\ \citenamefont {Coles}}]{melanson2025thermodynamic}%
  \BibitemOpen
  \bibfield  {author} {\bibinfo {author} {\bibfnamefont {Denis}\ \bibnamefont
  {Melanson}}, \bibinfo {author} {\bibfnamefont {Mohammad}\ \bibnamefont
  {Abu~Khater}}, \bibinfo {author} {\bibfnamefont {Maxwell}\ \bibnamefont
  {Aifer}}, \bibinfo {author} {\bibfnamefont {Kaelan}\ \bibnamefont
  {Donatella}}, \bibinfo {author} {\bibfnamefont {Max}\ \bibnamefont
  {Hunter~Gordon}}, \bibinfo {author} {\bibfnamefont {Thomas}\ \bibnamefont
  {Ahle}}, \bibinfo {author} {\bibfnamefont {Gavin}\ \bibnamefont {Crooks}},
  \bibinfo {author} {\bibfnamefont {Antonio~J}\ \bibnamefont {Martinez}},
  \bibinfo {author} {\bibfnamefont {Faris}\ \bibnamefont {Sbahi}}, \ and\
  \bibinfo {author} {\bibfnamefont {Patrick~J}\ \bibnamefont {Coles}},\
  }\bibfield  {title} {\enquote {\bibinfo {title} {Thermodynamic computing
  system for {AI} applications},}\ }\href@noop {} {\bibfield  {journal}
  {\bibinfo  {journal} {Nature Communications}\ }\textbf {\bibinfo {volume}
  {16}},\ \bibinfo {pages} {3757} (\bibinfo {year} {2025})}\BibitemShut
  {NoStop}%
\bibitem [{\citenamefont {Hagan}\ \emph {et~al.}(2011)\citenamefont {Hagan},
  \citenamefont {Elrad},\ and\ \citenamefont {Jack}}]{hagan2011mechanisms}%
  \BibitemOpen
  \bibfield  {author} {\bibinfo {author} {\bibfnamefont {Michael~F}\
  \bibnamefont {Hagan}}, \bibinfo {author} {\bibfnamefont {Oren~M}\
  \bibnamefont {Elrad}}, \ and\ \bibinfo {author} {\bibfnamefont {Robert~L}\
  \bibnamefont {Jack}},\ }\bibfield  {title} {\enquote {\bibinfo {title}
  {Mechanisms of kinetic trapping in self-assembly and phase transformation},}\
  }\href@noop {} {\bibfield  {journal} {\bibinfo  {journal} {The Journal of
  Chemical Physics}\ }\textbf {\bibinfo {volume} {135}} (\bibinfo {year}
  {2011})}\BibitemShut {NoStop}%
\bibitem [{\citenamefont {Whitelam}\ and\ \citenamefont
  {Jack}(2015)}]{whitelam2015statistical}%
  \BibitemOpen
  \bibfield  {author} {\bibinfo {author} {\bibfnamefont {Stephen}\ \bibnamefont
  {Whitelam}}\ and\ \bibinfo {author} {\bibfnamefont {Robert~L}\ \bibnamefont
  {Jack}},\ }\bibfield  {title} {\enquote {\bibinfo {title} {The statistical
  mechanics of dynamic pathways to self-assembly},}\ }\href@noop {} {\bibfield
  {journal} {\bibinfo  {journal} {Annual review of Physical Chemistry}\
  }\textbf {\bibinfo {volume} {66}},\ \bibinfo {pages} {143--163} (\bibinfo
  {year} {2015})}\BibitemShut {NoStop}%
\bibitem [{\citenamefont {Biroli}\ and\ \citenamefont
  {Garrahan}(2013)}]{biroli2013perspective}%
  \BibitemOpen
  \bibfield  {author} {\bibinfo {author} {\bibfnamefont {Giulio}\ \bibnamefont
  {Biroli}}\ and\ \bibinfo {author} {\bibfnamefont {Juan~P}\ \bibnamefont
  {Garrahan}},\ }\bibfield  {title} {\enquote {\bibinfo {title} {Perspective:
  The glass transition},}\ }\href@noop {} {\bibfield  {journal} {\bibinfo
  {journal} {The Journal of Chemical Physics}\ }\textbf {\bibinfo {volume}
  {138}} (\bibinfo {year} {2013})}\BibitemShut {NoStop}%
\bibitem [{\citenamefont {Arceri}\ \emph {et~al.}(2022)\citenamefont {Arceri},
  \citenamefont {Landes}, \citenamefont {Berthier},\ and\ \citenamefont
  {Biroli}}]{arceri2022glasses}%
  \BibitemOpen
  \bibfield  {author} {\bibinfo {author} {\bibfnamefont {Francesco}\
  \bibnamefont {Arceri}}, \bibinfo {author} {\bibfnamefont {Francois~P}\
  \bibnamefont {Landes}}, \bibinfo {author} {\bibfnamefont {Ludovic}\
  \bibnamefont {Berthier}}, \ and\ \bibinfo {author} {\bibfnamefont {Giulio}\
  \bibnamefont {Biroli}},\ }\bibfield  {title} {\enquote {\bibinfo {title}
  {Glasses and aging, a statistical mechanics perspective on},}\ }in\
  \href@noop {} {\emph {\bibinfo {booktitle} {Statistical and Nonlinear
  Physics}}}\ (\bibinfo  {publisher} {Springer},\ \bibinfo {year} {2022})\ pp.\
  \bibinfo {pages} {229--296}\BibitemShut {NoStop}%
\bibitem [{\citenamefont {M{\'e}zard}\ \emph {et~al.}(1984)\citenamefont
  {M{\'e}zard}, \citenamefont {Parisi}, \citenamefont {Sourlas}, \citenamefont
  {Toulouse},\ and\ \citenamefont {Virasoro}}]{mezard1984nature}%
  \BibitemOpen
  \bibfield  {author} {\bibinfo {author} {\bibfnamefont {Marc}\ \bibnamefont
  {M{\'e}zard}}, \bibinfo {author} {\bibfnamefont {Giorgio}\ \bibnamefont
  {Parisi}}, \bibinfo {author} {\bibfnamefont {Nicolas}\ \bibnamefont
  {Sourlas}}, \bibinfo {author} {\bibfnamefont {G{\'e}rard}\ \bibnamefont
  {Toulouse}}, \ and\ \bibinfo {author} {\bibfnamefont {Miguel}\ \bibnamefont
  {Virasoro}},\ }\bibfield  {title} {\enquote {\bibinfo {title} {Nature of the
  spin-glass phase},}\ }\href@noop {} {\bibfield  {journal} {\bibinfo
  {journal} {Physical Review Letters}\ }\textbf {\bibinfo {volume} {52}},\
  \bibinfo {pages} {1156} (\bibinfo {year} {1984})}\BibitemShut {NoStop}%
\bibitem [{\citenamefont {Ray}\ and\ \citenamefont
  {Crutchfield}(2023)}]{ray2023gigahertz}%
  \BibitemOpen
  \bibfield  {author} {\bibinfo {author} {\bibfnamefont {Kyle~J}\ \bibnamefont
  {Ray}}\ and\ \bibinfo {author} {\bibfnamefont {James~P}\ \bibnamefont
  {Crutchfield}},\ }\bibfield  {title} {\enquote {\bibinfo {title} {Gigahertz
  sub-{L}andauer momentum computing},}\ }\href@noop {} {\bibfield  {journal}
  {\bibinfo  {journal} {Physical Review Applied}\ }\textbf {\bibinfo {volume}
  {19}},\ \bibinfo {pages} {014049} (\bibinfo {year} {2023})}\BibitemShut
  {NoStop}%
\bibitem [{\citenamefont {Chupeau}\ \emph {et~al.}(2018)\citenamefont
  {Chupeau}, \citenamefont {Besga}, \citenamefont {Guery-Odelin}, \citenamefont
  {Trizac}, \citenamefont {Petrosyan},\ and\ \citenamefont
  {Ciliberto}}]{chupeau2018thermal}%
  \BibitemOpen
  \bibfield  {author} {\bibinfo {author} {\bibfnamefont {M.}~\bibnamefont
  {Chupeau}}, \bibinfo {author} {\bibfnamefont {B.}~\bibnamefont {Besga}},
  \bibinfo {author} {\bibfnamefont {D.}~\bibnamefont {Guery-Odelin}}, \bibinfo
  {author} {\bibfnamefont {E.}~\bibnamefont {Trizac}}, \bibinfo {author}
  {\bibfnamefont {A.}~\bibnamefont {Petrosyan}}, \ and\ \bibinfo {author}
  {\bibfnamefont {S.}~\bibnamefont {Ciliberto}},\ }\bibfield  {title} {\enquote
  {\bibinfo {title} {Thermal bath engineering for swift equilibration},}\
  }\href {\doibase 10.1103/PhysRevE.98.010104} {\bibfield  {journal} {\bibinfo
  {journal} {Physical Review E}\ }\textbf {\bibinfo {volume} {98}},\ \bibinfo
  {pages} {010104} (\bibinfo {year} {2018})}\BibitemShut {NoStop}%
\bibitem [{\citenamefont {Saha}\ \emph {et~al.}(2023)\citenamefont {Saha},
  \citenamefont {Ehrich}, \citenamefont {Gavrilov}, \citenamefont {Still},
  \citenamefont {Sivak},\ and\ \citenamefont
  {Bechhoefer}}]{saha2023information}%
  \BibitemOpen
  \bibfield  {author} {\bibinfo {author} {\bibfnamefont {Tushar~K.}\
  \bibnamefont {Saha}}, \bibinfo {author} {\bibfnamefont {Jannik}\ \bibnamefont
  {Ehrich}}, \bibinfo {author} {\bibfnamefont {Mom{\v{c}}ilo}\ \bibnamefont
  {Gavrilov}}, \bibinfo {author} {\bibfnamefont {Susanne}\ \bibnamefont
  {Still}}, \bibinfo {author} {\bibfnamefont {David~A.}\ \bibnamefont {Sivak}},
  \ and\ \bibinfo {author} {\bibfnamefont {John}\ \bibnamefont {Bechhoefer}},\
  }\bibfield  {title} {\enquote {\bibinfo {title} {Information engine in a
  nonequilibrium bath},}\ }\href {\doibase 10.1103/PhysRevLett.131.057101}
  {\bibfield  {journal} {\bibinfo  {journal} {Physical Review Letters}\
  }\textbf {\bibinfo {volume} {131}},\ \bibinfo {pages} {057101} (\bibinfo
  {year} {2023})}\BibitemShut {NoStop}%
\bibitem [{Note1()}]{Note1}%
  \BibitemOpen
  \bibinfo {note} {A similar effect could be achieved by rescaling $V \to
  \lambda V$ and increasing temperature $T \to \lambda T$, but we assume
  scaling temperature is less practical than adding an external source of
  noise.}\BibitemShut {Stop}%
\bibitem [{\citenamefont {Whitelam}\ and\ \citenamefont
  {Casert}(2024)}]{whitelam2024thermodynamic}%
  \BibitemOpen
  \bibfield  {author} {\bibinfo {author} {\bibfnamefont {Stephen}\ \bibnamefont
  {Whitelam}}\ and\ \bibinfo {author} {\bibfnamefont {Corneel}\ \bibnamefont
  {Casert}},\ }\bibfield  {title} {\enquote {\bibinfo {title} {Thermodynamic
  computing out of equilibrium},}\ }\href@noop {} {\bibfield  {journal}
  {\bibinfo  {journal} {arXiv preprint arXiv:2412.17183}\ } (\bibinfo {year}
  {2024})}\BibitemShut {NoStop}%
\bibitem [{\citenamefont {Gardiner}(2009)}]{gardiner2009stochastic}%
  \BibitemOpen
  \bibfield  {author} {\bibinfo {author} {\bibfnamefont {Crispin~W.}\
  \bibnamefont {Gardiner}},\ }\href@noop {} {\emph {\bibinfo {title}
  {Stochastic Methods: A Handbook for the Natural and Social Sciences}}},\
  \bibinfo {edition} {4th}\ ed.\ (\bibinfo  {publisher} {Springer},\ \bibinfo
  {year} {2009})\BibitemShut {NoStop}%
\bibitem [{Note2()}]{Note2}%
  \BibitemOpen
  \bibinfo {note} {For ease of display the matrix elements are shown to not
  more than 6 significant figures.}\BibitemShut {Stop}%
\bibitem [{\citenamefont {Hooker}(2021)}]{hooker2021hardware}%
  \BibitemOpen
  \bibfield  {author} {\bibinfo {author} {\bibfnamefont {Sara}\ \bibnamefont
  {Hooker}},\ }\bibfield  {title} {\enquote {\bibinfo {title} {The hardware
  lottery},}\ }\href@noop {} {\bibfield  {journal} {\bibinfo  {journal}
  {Communications of the ACM}\ }\textbf {\bibinfo {volume} {64}},\ \bibinfo
  {pages} {58--65} (\bibinfo {year} {2021})}\BibitemShut {NoStop}%
\bibitem [{\citenamefont {Preskill}(2018)}]{preskill2018quantum}%
  \BibitemOpen
  \bibfield  {author} {\bibinfo {author} {\bibfnamefont {John}\ \bibnamefont
  {Preskill}},\ }\bibfield  {title} {\enquote {\bibinfo {title} {Quantum
  computing in the {NISQ} era and beyond},}\ }\href@noop {} {\bibfield
  {journal} {\bibinfo  {journal} {Quantum}\ }\textbf {\bibinfo {volume} {2}},\
  \bibinfo {pages} {79} (\bibinfo {year} {2018})}\BibitemShut {NoStop}%
\bibitem [{\citenamefont {Sekimoto}(1998)}]{sekimoto1998langevin}%
  \BibitemOpen
  \bibfield  {author} {\bibinfo {author} {\bibfnamefont {Ken}\ \bibnamefont
  {Sekimoto}},\ }\bibfield  {title} {\enquote {\bibinfo {title} {Langevin
  equation and thermodynamics},}\ }\href@noop {} {\bibfield  {journal}
  {\bibinfo  {journal} {Progress of Theoretical Physics Supplement}\ }\textbf
  {\bibinfo {volume} {130}},\ \bibinfo {pages} {17--27} (\bibinfo {year}
  {1998})}\BibitemShut {NoStop}%
\bibitem [{\citenamefont {Seifert}(2012)}]{seifert2012stochastic}%
  \BibitemOpen
  \bibfield  {author} {\bibinfo {author} {\bibfnamefont {Udo}\ \bibnamefont
  {Seifert}},\ }\bibfield  {title} {\enquote {\bibinfo {title} {Stochastic
  thermodynamics, fluctuation theorems and molecular machines},}\ }\href@noop
  {} {\bibfield  {journal} {\bibinfo  {journal} {Reports on Progress in
  Physics}\ }\textbf {\bibinfo {volume} {75}},\ \bibinfo {pages} {126001}
  (\bibinfo {year} {2012})}\BibitemShut {NoStop}%
\bibitem [{\citenamefont {Horowitz}\ and\ \citenamefont
  {Hill}(2015)}]{Horowitz2015}%
  \BibitemOpen
  \bibfield  {author} {\bibinfo {author} {\bibfnamefont {Paul}\ \bibnamefont
  {Horowitz}}\ and\ \bibinfo {author} {\bibfnamefont {Winfield}\ \bibnamefont
  {Hill}},\ }\href@noop {} {\emph {\bibinfo {title} {The Art of
  Electronics}}},\ \bibinfo {edition} {3rd}\ ed.\ (\bibinfo  {publisher}
  {Cambridge University Press},\ \bibinfo {address} {Cambridge, UK},\ \bibinfo
  {year} {2015})\BibitemShut {NoStop}%
\bibitem [{\citenamefont {Falc{\'o}n}\ and\ \citenamefont
  {Falcon}(2009)}]{falcon2009fluctuations}%
  \BibitemOpen
  \bibfield  {author} {\bibinfo {author} {\bibfnamefont {Claudio}\ \bibnamefont
  {Falc{\'o}n}}\ and\ \bibinfo {author} {\bibfnamefont {Eric}\ \bibnamefont
  {Falcon}},\ }\bibfield  {title} {\enquote {\bibinfo {title} {Fluctuations of
  energy flux in a simple dissipative out-of-equilibrium system},}\ }\href@noop
  {} {\bibfield  {journal} {\bibinfo  {journal} {Physical Review E:
  Statistical, Nonlinear, and Soft Matter Physics}\ }\textbf {\bibinfo {volume}
  {79}},\ \bibinfo {pages} {041110} (\bibinfo {year} {2009})}\BibitemShut
  {NoStop}%
\bibitem [{\citenamefont {Gomez-Solano}\ \emph {et~al.}(2010)\citenamefont
  {Gomez-Solano}, \citenamefont {Bellon}, \citenamefont {Petrosyan},\ and\
  \citenamefont {Ciliberto}}]{gomez2010steady}%
  \BibitemOpen
  \bibfield  {author} {\bibinfo {author} {\bibfnamefont {Juan~Ruben}\
  \bibnamefont {Gomez-Solano}}, \bibinfo {author} {\bibfnamefont {Ludovic}\
  \bibnamefont {Bellon}}, \bibinfo {author} {\bibfnamefont {Artyom}\
  \bibnamefont {Petrosyan}}, \ and\ \bibinfo {author} {\bibfnamefont {Sergio}\
  \bibnamefont {Ciliberto}},\ }\bibfield  {title} {\enquote {\bibinfo {title}
  {Steady-state fluctuation relations for systems driven by an external random
  force},}\ }\href@noop {} {\bibfield  {journal} {\bibinfo  {journal}
  {Europhysics Letters}\ }\textbf {\bibinfo {volume} {89}},\ \bibinfo {pages}
  {60003} (\bibinfo {year} {2010})}\BibitemShut {NoStop}%
\bibitem [{\citenamefont {Esposito}(2012)}]{esposito2012stochastic}%
  \BibitemOpen
  \bibfield  {author} {\bibinfo {author} {\bibfnamefont {Massimiliano}\
  \bibnamefont {Esposito}},\ }\bibfield  {title} {\enquote {\bibinfo {title}
  {Stochastic thermodynamics under coarse graining},}\ }\href@noop {}
  {\bibfield  {journal} {\bibinfo  {journal} {Physical Review E -- Statistical,
  Nonlinear, and Soft Matter Physics}\ }\textbf {\bibinfo {volume} {85}},\
  \bibinfo {pages} {041125} (\bibinfo {year} {2012})}\BibitemShut {NoStop}%
\bibitem [{\citenamefont {Shiraishi}\ and\ \citenamefont
  {Sagawa}(2015)}]{shiraishi2015fluctuation}%
  \BibitemOpen
  \bibfield  {author} {\bibinfo {author} {\bibfnamefont {Naoto}\ \bibnamefont
  {Shiraishi}}\ and\ \bibinfo {author} {\bibfnamefont {Takahiro}\ \bibnamefont
  {Sagawa}},\ }\bibfield  {title} {\enquote {\bibinfo {title} {Fluctuation
  theorem for partially masked nonequilibrium dynamics},}\ }\href@noop {}
  {\bibfield  {journal} {\bibinfo  {journal} {Physical Review E}\ }\textbf
  {\bibinfo {volume} {91}},\ \bibinfo {pages} {012130} (\bibinfo {year}
  {2015})}\BibitemShut {NoStop}%
\bibitem [{\citenamefont {Fodor}\ \emph {et~al.}(2016)\citenamefont {Fodor},
  \citenamefont {Nardini}, \citenamefont {Cates}, \citenamefont {Tailleur},
  \citenamefont {Visco},\ and\ \citenamefont {Van~Wijland}}]{fodor2016how}%
  \BibitemOpen
  \bibfield  {author} {\bibinfo {author} {\bibfnamefont {{\'E}tienne}\
  \bibnamefont {Fodor}}, \bibinfo {author} {\bibfnamefont {Cesare}\
  \bibnamefont {Nardini}}, \bibinfo {author} {\bibfnamefont {Michael~E}\
  \bibnamefont {Cates}}, \bibinfo {author} {\bibfnamefont {Julien}\
  \bibnamefont {Tailleur}}, \bibinfo {author} {\bibfnamefont {Paolo}\
  \bibnamefont {Visco}}, \ and\ \bibinfo {author} {\bibfnamefont
  {Fr{\'e}d{\'e}ric}\ \bibnamefont {Van~Wijland}},\ }\bibfield  {title}
  {\enquote {\bibinfo {title} {How far from equilibrium is active matter?}}\
  }\href@noop {} {\bibfield  {journal} {\bibinfo  {journal} {Physical Review
  Letters}\ }\textbf {\bibinfo {volume} {117}},\ \bibinfo {pages} {038103}
  (\bibinfo {year} {2016})}\BibitemShut {NoStop}%
\bibitem [{\citenamefont {Marconi}\ \emph {et~al.}(2017)\citenamefont
  {Marconi}, \citenamefont {Puglisi},\ and\ \citenamefont
  {Maggi}}]{marconi2017heat}%
  \BibitemOpen
  \bibfield  {author} {\bibinfo {author} {\bibfnamefont {Umberto
  Marini~Bettolo}\ \bibnamefont {Marconi}}, \bibinfo {author} {\bibfnamefont
  {Andrea}\ \bibnamefont {Puglisi}}, \ and\ \bibinfo {author} {\bibfnamefont
  {Claudio}\ \bibnamefont {Maggi}},\ }\bibfield  {title} {\enquote {\bibinfo
  {title} {Heat, temperature and clausius inequality in a model for active
  brownian particles},}\ }\href@noop {} {\bibfield  {journal} {\bibinfo
  {journal} {Scientific reports}\ }\textbf {\bibinfo {volume} {7}},\ \bibinfo
  {pages} {46496} (\bibinfo {year} {2017})}\BibitemShut {NoStop}%
\bibitem [{\citenamefont {Puglisi}\ \emph {et~al.}(2017)\citenamefont
  {Puglisi}, \citenamefont {Sarracino},\ and\ \citenamefont
  {Vulpiani}}]{puglisi2017temperature}%
  \BibitemOpen
  \bibfield  {author} {\bibinfo {author} {\bibfnamefont {Andrea}\ \bibnamefont
  {Puglisi}}, \bibinfo {author} {\bibfnamefont {Alessandro}\ \bibnamefont
  {Sarracino}}, \ and\ \bibinfo {author} {\bibfnamefont {Angelo}\ \bibnamefont
  {Vulpiani}},\ }\bibfield  {title} {\enquote {\bibinfo {title} {Temperature in
  and out of equilibrium: A review of concepts, tools and attempts},}\
  }\href@noop {} {\bibfield  {journal} {\bibinfo  {journal} {Physics Reports}\
  }\textbf {\bibinfo {volume} {709}},\ \bibinfo {pages} {1--60} (\bibinfo
  {year} {2017})}\BibitemShut {NoStop}%
\bibitem [{\citenamefont {Mart�nez}\ \emph {et~al.}(2012)\citenamefont
  {Martinez}, \citenamefont {Roldan}, \citenamefont {Parrondo},\ and\
  \citenamefont {Petrov}}]{martinez2012effective}%
  \BibitemOpen
  \bibfield  {author} {\bibinfo {author} {\bibfnamefont {I.~A.}\ \bibnamefont
  {Martinez}}, \bibinfo {author} {\bibfnamefont {E.}~\bibnamefont {Roldan}},
  \bibinfo {author} {\bibfnamefont {J.~M.~R.}\ \bibnamefont {Parrondo}}, \ and\
  \bibinfo {author} {\bibfnamefont {D.}~\bibnamefont {Petrov}},\ }\bibfield
  {title} {\enquote {\bibinfo {title} {Effective heating to several thousand
  {K}elvins of an optically trapped sphere in a liquid},}\ }\href {\doibase
  10.1103/PhysRevE.87.032159} {\bibfield  {journal} {\bibinfo  {journal}
  {Physical Review E}\ }\textbf {\bibinfo {volume} {87}},\ \bibinfo {pages}
  {032159} (\bibinfo {year} {2012})}\BibitemShut {NoStop}%
\bibitem [{\citenamefont {Ciliberto}(2017)}]{Ciliberto2017experiments}%
  \BibitemOpen
  \bibfield  {author} {\bibinfo {author} {\bibfnamefont {Sergio}\ \bibnamefont
  {Ciliberto}},\ }\bibfield  {title} {\enquote {\bibinfo {title} {Experiments
  in stochastic thermodynamics: Short history and perspectives},}\ }\href@noop
  {} {\bibfield  {journal} {\bibinfo  {journal} {Physical Review X}\ }\textbf
  {\bibinfo {volume} {7}},\ \bibinfo {pages} {021051} (\bibinfo {year}
  {2017})}\BibitemShut {NoStop}%
\bibitem [{Note3()}]{Note3}%
  \BibitemOpen
  \bibinfo {note} {Explicitly, we can show from Eq.~(\ref {action}) that the
  heat flow vanishes, $q[x(t)]=\protect \mathaccentV {tilde}07E{T}[\protect
  \mathcal {A}(x(t),0) - \protect \mathcal {A}(x(t_{\protect \rm f}-t),0)]=0$,
  and so therefore does the entropy production.}\BibitemShut {Stop}%
\end{thebibliography}
\end{document}